\documentclass[twocolumn, superscriptaddress,preprintnumbers,amsmath,amssymb,longbibliography]{revtex4-1}
\usepackage{graphicx}
\usepackage{dcolumn}
\usepackage{bm}
\usepackage{mathtools}
\usepackage{epstopdf}
\usepackage{color,soul}

\begin{document}

\preprint{Version: \today}

\title{Realization and modeling of rf superconducting quantum interference device metamaterials}

\author{M. Trepanier}
\email{mctrep@umd.edu}
\thanks{these authors contributed equally}
\affiliation{Department of Physics, CNAM, University of Maryland, College Park, Maryland 20742-4111, USA}
\author{Daimeng Zhang}
\email{dmchang@umd.edu}
\thanks{these authors contributed equally}
\affiliation{Department of Electrical and Computer Engineering, University of Maryland, College Park, Maryland 20742-3285, USA}
\author{Oleg Mukhanov}
\affiliation{Hypres, Inc., 175 Clearbrook Road, Elmsford, New York 10523, USA}
\author{Steven M. Anlage}
\affiliation{Department of Physics, CNAM, University of Maryland, College Park, Maryland 20742-4111, USA}
\affiliation{Department of Electrical and Computer Engineering, University of Maryland, College Park, Maryland 20742-3285, USA}

\date{\today}
\begin{abstract}
We have prepared meta-atoms based on radio frequency superconducting quantum interference devices (RF SQUIDs) and examined their tunability with dc magnetic field, rf current, and temperature. RF SQUIDs are superconducting split ring resonators in which the usual capacitance is supplemented with a Josephson junction, which introduces strong nonlinearity in the rf properties. We find excellent agreement between the data and a model which regards the Josephson junction as the resistively and capacitively-shunted junction. A magnetic field tunability of 80 THz/Gauss at 12 GHz is observed, a total tunability of 56$\%$ is achieved, and a unique electromagnetically-induced transparency feature at intermediate excitation powers is demonstrated for the first time. An RF SQUID metamaterial is shown to have qualitatively the same behavior as a single RF SQUID with regards to DC flux and temperature tuning.
\end{abstract}

\maketitle

\section{Introduction}
Metamaterials are artificially structured media designed to have electromagnetic properties not found in nature.  These properties arise from both the structure of individual meta-atoms and the interactions between them, resulting in interesting collective behavior.  A wide variety of properties and applications have been pursued using metamaterials, including negative index of refraction \cite{Veselago1968} \cite{Smith2000} \cite{Shelby2001}, super-resolution imaging \cite{Pendry2000} \cite{Jacob2006}, cloaking \cite{Alu2003} \cite{Schurig2006}, transformation optics \cite{Leonhardt2006} \cite{Pendry2006}, and perfect absorption \cite{Landy2008}.

Most of the applications invoking novel optical properties impose three stringent constraints.  First, the metamaterial must have low attenuation.  Features such as evanescent wave amplification \cite{Liu2003} and negative refraction are strongly suppressed by even small amounts of loss \cite{Ruppin2000} \cite{Ruppin2001} \cite{Haldane2002} \cite{Smith2003} \cite{Liu2003} \cite{Koschny2006}.  For example, the enhanced loss in metamaterials approaching the plasmonic limit has imposed a severe limitation on visible wavelength metamaterials composed of noble metal nano-structures \cite{Dimmock2003} \cite{Zhou2005} \cite{Ishikawa2005} \cite{Urzhumov2008}. Secondly, the meta-atoms must be of deep sub-wavelength dimensions to achieve the metamaterial limit, as opposed to the photonic crystal limit.  This has been an issue both in the visible and microwave regimes, where meta-atom sizes often approach the scale of the wavelength to minimize losses \cite{Ziolkowski2003} \cite{Ziolkowski2006}.  Thirdly, it is desirable to make metamaterials that have textured properties in space (e.g. for cloaking or transformation optics), or that can be tuned and re-configured after the metamaterial has been fabricated \cite{Zharov2003}.  The ability to tune the electromagnetic response over a wide range, and on short time scales, is desirable for applications such as software-defined radio \cite{Wikborg1999} and filters for digital rf receivers \cite{Mukhanov2001} \cite{Mukhanov2008}.
Superconducting metamaterials have been proposed to address all three of these constraints \cite{Anlage2011}.  In addition, the quantum coherent nature of the superconducting state leads to qualitatively new phenomena such as fluxoid quantization and Josephson effects. Here we focus on the tunability and reconfigurable nature of superconducting Josephson metamaterials.

\subsection{Tunable Superconducting Metamaterials}
Superconductors are fundamentally nonlinear due to the nonlinear Meissner effect \cite{Gittleman1965} \cite{Yip1992}.
In general this intrinsic nonlinearity will be encountered near the limits of parameter space spanned by temperature, applied magnetic field, and applied current.

Superconducting metamaterials are tunable with temperature due to the temperature dependence of the superfluid density and kinetic inductance \cite{Ricci2005} \cite{Ricci2006}. This type of tunability has been exploited in various superconducting metamaterials with some success \cite{Gu2010} \cite{Fedotov2010} \cite{Chen2010} \cite{Wu2011} \cite{Kurter2012}. However, temperature tuning is slow since the thermal inertia of the meta-atoms can be large, even at low temperatures. Typical estimates for temperature tuning response times are on the order of 10 $\mu$s \cite{Savinov2012}.

Applied currents can also be used to tune superconducting metamaterials; they can cause superfluid de-pairing, which increases the kinetic inductance \cite{Anlage1989}. For example, applied currents can tune sub-THz transmission of a metamaterial composed of a network of resonators connected by a superconducting wire loop \cite{Savinov2012}. However, applied currents often induce magnetic vortices in the superconductor before the de-pairing critical current is reached \cite{Tahara1990} \cite{Savinov2012}. These vortices move under the influence of the high-frequency currents, creating enhanced inductance and dissipation. The additional dissipation renders the superconductors less attractive for applications.

Magnetic field tuning of superconductors is attractive for applications because it can have a large effect on metamaterial properties. For example, superconducting split-ring resonators (SRRs) have been tuned by both DC and RF magnetic fields \cite{Ricci2007} \cite{Jin2010}. A DC magnetic field was shown to add magnetic flux to a thin film Nb SRR, increasing its inductance and loss \cite{Ricci2007}.  The RF magnetic field created enhanced RF screening currents at discrete locations in the SRR, resulting in enhanced inductance and dissipation as magnetic flux moved into and out of the superconducting film at high frequency \cite{Ricci2007} \cite{Zhuravel20062} \cite{Zhuravel20063} \cite{Kurter20111} \cite{Kurter20112} \cite{Zhuravel2012}. It was also found that a superconducting resonator exhibiting an analog of electromagnetically-induced transparency showed a strong switching behavior at high excitation power \cite{Kurter2012}, for similar reasons. However, the insertion of magnetic flux into superconducting materials is often too slow and too dissipative for tuning applications; a low-dissipation quantum effect would be better. For example, flux quantization could be used to discretely tune a superconducting meta-atom \cite{Savinov20122}.The addition of the Josephson effect to superconducting metamaterials adds a mechanism of tunability that offers a large degree of high speed tunability and low dissipation.

\subsection{Superconducting Meta-Atoms Employing the Josephson Effect}
A superconductor can be described by a macroscopic phase-coherent complex quantum wavefunction $\Psi=\sqrt{n_s}e^{i \theta}$ \cite{Tinkham1996}.  This wavefunction inherits its phase coherence from the underlying microscopic BCS (Bardeen, Cooper, Schrieffer) wavefunction describing the Cooper pairing of electrons in the metal. The absolute square of the macroscopic quantum wavefunction is the local superfluid density $n_s \sim |\Psi|^2$ \cite{Orlando1991}.  When two superconductors are brought close together and separated by a thin insulating barrier, there can be tunneling of Cooper pairs between the two materials \cite{Likharev1986} \cite{Tinkham1996}. This tunneling results in two types of Josephson effect.

The DC Josephson effect produces a DC current between the two superconductors which depends on the phase difference between their macroscopic quantum wavefunctions as $I=I_c \sin{\delta(t)}$, where $\delta(t)=\theta_1(t)-\theta_2(t)-\frac{2 \pi}{\Phi_0} \int_1^2 \vec{A}(\vec{r},t) \cdot d\vec{l}$ is the gauge-invariant phase difference between superconductors 1 and 2, $\vec{A}(\vec{r},t)$ is the magnetic vector potential in the region between the superconductors,  $I_c$ is the critical (maximum) current of the junction, and $\Phi_0=\frac{h}{2e} \cong 2.07 \times 10^{-15}$ Tm$^2$ is the flux quantum ($h$ is Planck's constant and $e$ is the electronic charge).

The AC Josephson effect relates a DC voltage drop across the junction to a time-varying gauge-invariant phase difference, and therefore to an AC current in the junction: $\frac{d \delta}{dt}=\frac{2 \pi}{\Phi_0}V$, where $V$ is the DC voltage across the junction. In general the AC impedance of a Josephson junction contains both resistive and reactive components \cite{Auracher1973}.

In the presence of both DC and AC currents, the inductance of the junction can be described approximately as \cite{Orlando1991}
\begin{equation}
\label{Ljj}
L_{JJ}=\frac{\Phi_0}{2 \pi I_c \cos{\delta}}.
\end{equation}
The Josephson inductance is a strong function of applied currents and fields. Tuning of Josephson inductance in a transmission line metamaterial geometry was considered theoretically by Salehi, Majedi, and Mansour \cite{Salehi2005} \cite{Salehi2007}.

\subsection{SQUID Metamaterials}
\begin{figure}[]
\includegraphics*[width=80mm]{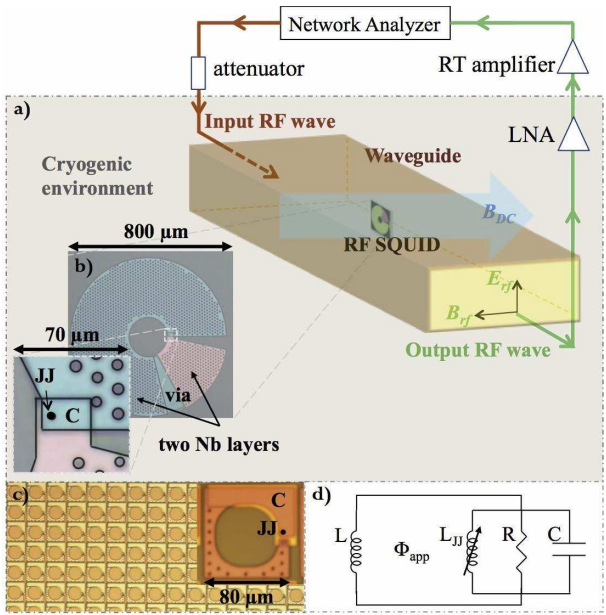}
\caption{a) Schematic diagram of the experiment showing the flow of rf signal from the network analyzer (at room temperature, RT), through the attenuator, into the waveguide (in the cryogenic environment), through the RF SQUID biased by DC magnetic field, out of the waveguide through a series of amplifiers, and back to the network analyzer. b) Micrograph of an RF SQUID meta-atom, consisting of two perforated Nb layers connected by a via and Josephson junction (JJ). The overlap capacitance is the small rectangular region immediately surrounding the junction. c) Micrograph of a portion of the 27x27 RF SQUID array and a single meta-atom that composes the array. d) Circuit diagram of the RF SQUID modeled as an RCSJ in parallel with loop inductance $L$.}
\label{figure1}
\end{figure}

In this work we focus on meta-atoms comprised of a superconducting loop interrupted by a single Josephson junction, commonly known as a radio frequency superconducting quantum interference device (RF SQUID). Since the RF SQUID meta-atom (shown in Fig. \ref{figure1}(b))
is a natural quantum analog of the SRR, we examine the electromagnetic properties of the SQUID near its self-resonant frequency.

The original purpose of the RF SQUID was to measure small magnetic fields and operate as a flux to frequency transducer \cite{Silver1967} \cite{Chesca2001}. The first proposal to use an array of RF SQUIDs as a metamaterial was made by Du, Chen, and Li \cite{Du2006} \cite{Du2008}. Their calculation assumes the SQUID has quantized energy levels and considers the interaction of microwave photons with the lowest states of the SQUID potential.  For small detuning of the photon frequency above the transition from the ground state to the first excited state, the medium presents a negative effective permeability. The frequency region of negative permeability is diminished by a non-zero de-phasing rate and negative permeability will disappear for de-phasing rates larger than a critical value.

RF SQUIDs interacting with classical electromagnetic fields were modelled by Lazarides and Tsironis \cite{Lazarides2007}.  They considered a two-dimensional array of RF SQUIDs in which the Josephson junction was treated as a parallel combination of resistance, capacitance, and Josephson inductance.  Near resonance, a single RF SQUID can have a large diamagnetic response. An RF SQUID array displays a negative real part of effective permeability for a range of a frequencies and magnetic fields. The permeability is oscillatory as a function of applied magnetic flux, and is suppressed with applied fields that induce currents in excess of the critical current of the Josephson junction. Similar calculations, which included interactions between the RF SQUIDs, were carried out by Maimistov and Gabitov \cite{Maimistov2010}. The response of a two-dimensional RF SQUID metamaterial under an applied oscilating magnetic field was also numerically investigated by Lazarides and Tsironis \cite{Lazarides2013}. The weakly coupled elements of the metamaterial are predicted to show bistability and synchronization.

Related work on a one-dimensional array of superconducting islands that can act as quantum bits (qubits) was considered by Rakhmanov, et al. \cite{Rakhmanov2008}.  When interacting with classical electromagnetic radiation, the array can create a quantum photonic crystal, which can support a variety of nonlinear wave excitations.  A similar idea based on a SQUID transmission line was implemented to perform parametric amplification of microwave signals \cite{Castellanos2007} \cite{Beltran2008}. A one-dimensional array of dc SQUIDs has been utilized as a tunable composite right/left-handed transmission line \cite{Ovchinnikova2013}. A dc-field tunable one-dimensional SQUID metamaterial embedded in a co-planar waveguide structure has been recently demonstrated \cite{Jung2013} \cite{Butz20131} \cite{Butz20132}.

It is expected that SQUID metamaterials will be able to perform functions similar to galvanically connected SQUID arrays which have been developed in recent years.  One possible application takes advantage of the SQUID's extreme sensitivity to magnetic flux to create compact, wideband antennas, sensitive to high frequency magnetic fields, as opposed to electric fields \cite{Luine1999} \cite{Caputo2007} \cite{Kornev2012} \cite{Kornev2013}.
Other possible applications include low noise amplifiers for RF sensing and qubit readout \cite{Snigirev2007} \cite{Kornev2009} \cite{Beltran2008}, and highly sensitive magnetometers and filters \cite{Jeng2005} \cite{Bruno2003} \cite{Oppenlander2003}.

Our objective in this paper is to demonstrate the first SQUID metamaterial in a geometry that can be easily extended to a fully three-dimensional structure which can interact with free-space electromagnetic waves.  We demonstrate that RF SQUID meta-atoms are remarkably tunable with very small DC magnetic fields, as well as temperature and RF currents.

The remainder of the paper is organized as follows. We first present models of RF SQUID linear and nonlinear behavior. Then we discuss the design and fabrication of a single RF SQUID meta-atom and a dense array of such SQUIDs.  Section \ref{results} presents our experimental results of the RF SQUID meta-atom and array in waveguide geometries, demonstrating tunability of the meta-atom properties with DC flux, temperature, and RF flux.  Section \ref{discussion} is a comparison of data and model predictions, and discussion of the results, and Section \ref{conclusion} serves as a conclusion.

\section{Modeling}
\label{model}
An ideal Josephson junction can be treated as an inductor with a variable inductance given by Eq. \ref{Ljj}. A more realistic model (valid for low frequencies and currents) shunts the junction with a sub-gap resistance $R$ (representing the tunneling of normal state electrons across the junction) and capacitance $C$ (the capacitance of two overlapping conductors separated by an insulator) \cite{Orlando1991}. We model the RF SQUID as a resistively and capacitively shunted junction (RCSJ) in parallel with an inductor representing the inductance, $L$, of the superconducting loop. The result is an RLC circuit (see circuit diagram in Fig. \ref{figure1}(d)) with a resonant frequency given by
\begin{equation}
\label{resfreq}
f_0(T,\Phi_{app})=\frac{1}{2 \pi\sqrt{\left(\frac{1}{L}+\frac{1}{L_{JJ}(T,\Phi_{app})}\right)^{-1}C}},
\end{equation}
where $T$ is the temperature and $\Phi_{app}$ is the magnetic flux applied to the SQUID loop. The quality factor for this parallel RLC circuit is proportional to the resonant frequency and given by,
\begin{equation}
\label{Q}
Q=R\sqrt{\frac{C}{\left(\frac{1}{L}+\frac{1}{L_{JJ}(T,\Phi_{app})}\right)^{-1}}}=2\pi RC f_0(T,\Phi_{app}).
\end{equation}
To understand the behavior of the resonance as a function of temperature and applied flux it is necessary to solve for the time-dependent gauge-invariant phase difference $\delta(T, \Phi_{app})$.

We use the RCSJ model to solve for $\delta$ subject to the condition that the total flux through the loop must be an integer number ($n$) of flux quanta, i.e. $\Phi=n \Phi_0$. The total flux, $\Phi$, is related to the applied flux, $\Phi_{app}$, and the current-induced flux by
\begin{equation}
\Phi_{app}= \Phi+LI
\label{totalflux}
\end{equation}
where $I$ is the current through the loop, which is the sum of the currents through the junction, the resistor, and the capacitor. We can rewrite Eq. \ref{totalflux} in the terms of $\delta$ as
\begin{equation}
\label{5}
\begin{multlined}
\Phi_{DC}+\Phi_{rf} \sin{\omega t}= \frac{\Phi_0 \delta}{2 \pi}+ \\
L\left(I_c \sin{\delta} + \frac{1}{R}\frac{\Phi_0}{2 \pi}\frac{d \delta}{dt}+C\frac{\Phi_0}{2 \pi}\frac{d^2 \delta}{dt^2}\right)
\end{multlined}
\end{equation}
where $\Phi_{DC}$ is the DC flux bias of the applied magnetic field and $\Phi_{rf}$ and $\omega$ are the amplitude and angular frequency of the applied rf field respectively. By making the substitution $\Phi=\frac{\Phi_0 \delta}{2 \pi}$, we have taken $n=1$ because the choice of integer shifts $\delta$ (which has $2 \pi$ periodicity) by $2 \pi$. Equation \ref{5} can be recast in non-dimensional form as
\begin{equation}
\label{nondim}
\begin{multlined}
2 \pi \left[f_{DC}+f_{rf} \sin{\Omega \tau}\right]= \\
\delta+\beta_{rf} \sin{\delta}+\frac{1}{Q_{geo}} \frac{d \delta}{d \tau}+\frac{d^2 \delta}{d\tau^2}
\end{multlined}
\end{equation}
where $\omega_{0,geo}=\sqrt{\frac{1}{LC}}$, $\tau=\omega_{0,geo} t$, $\Omega=\frac{\omega}{\omega_{0,geo}}$, $\beta_{rf}=\frac{2 \pi L I_c}{\Phi_0}$, $Q_{geo}=R\sqrt{\frac{C}{L}}$, $f_{rf}=\frac{\Phi_{rf}}{\Phi_0}$, and $f_{DC}=\frac{\Phi_{DC}}{\Phi_0}$.

A well-studied mechanical analogue of the Josephson junction is the driven and damped pendulum \cite{Falco1976}. Equation \ref{nondim} resembles the nonlinear differential equation that governs that system. However, incorporating the junction into a superconducting loop subject to flux quantization introduces an additional term linear in $\delta$ on the right-hand side of Eq. \ref{nondim}. Unlike the driven, damped pendulum, this equation is linear in the high power limit in addition to the low power limit. For our chosen parameters, $\delta$ is always periodic in time and its  variation is dominated by the same frequency as the driving flux. We do not observe chaos for parameter values relevant to the experiments discussed below.

At intermediate powers the full non-linear Eq. \ref{nondim} must be solved for $\delta$, and therefore $f_0(T,f_{DC},f_{rf})$, but the equation can be linearized and simplified in the low and high power limits. In the low rf power limit (where it is assumed that the time-varying component of the gauge invariant phase difference is very small i.e. $\delta_{rf}(\tau)<<1$) Eq. \ref{nondim} can be linearized by separating the phase difference into a DC and rf components, i.e.
\begin{equation}
\delta=\delta_{DC}+\delta_{rf}(\tau).
\end{equation}
Equation \ref{nondim} simplifies to the following time-independent and time-dependent equations:
\begin{equation}
2\pi f_{DC}=\delta_{DC}+\beta_{rf}\sin{\delta_{DC}}
\label{dconly}
\end{equation}
\begin{equation}
\label{lowrfdiff}
2 \pi f_{rf} \sin{\Omega \tau}= \alpha \delta_{rf}+\frac{1}{Q_{geo}}\frac{d \delta_{rf}}{d \tau}+\frac{d^2 \delta_{rf}}{d\tau^2}
\end{equation}
where $\alpha = 1+\beta_{rf}\cos{\delta_{DC}}$. In this case the DC flux dictates $\delta_{DC}$ and Eq. \ref{lowrfdiff} has an analytic solution for $\delta_{rf}$ given by
\begin{equation}
\label{lowrfsol}
\delta_{rf}=2 \pi f_{rf}\frac{ (\alpha - \Omega^2)\sin{\Omega \tau}-(\Omega/Q_{geo}) \cos{\Omega \tau}}{(\alpha -\Omega^2)^2+(\Omega/Q_{geo})^2}.
\end{equation}
From this result one finds that $\delta$ and consequently $f_0(T, f_{DC})$ are periodic functions of DC flux, $f_{DC}$, with maximum and minimum values at $f_{DC} = n$, and $f_{DC} = n+1/2$ respectively, where $n$ is any integer.

The resonant frequency of the SQUID, $f_0$, is also temperature dependent due to the temperature dependence of the critical current, $I_c$. As temperature increases the maximum frequency decreases and the minimum frequency increases resulting in a reduction of total flux tunability of the SQUID.
When the temperature reaches the critical temperature of niobium ($T_c=9.2$ K), $|L_{JJ}|$ is expected to diverge (since $I_c\rightarrow0$). In this case the Josephson junction loses the ability to modify the resonance and the resonant frequency reduces to
\begin{equation}
\label{geofreq}
f_{0,geo} = \frac{1}{2\pi \sqrt{LC}}.
\end{equation}
This resonance depends solely on the non-junction properties of the SQUID and is insensitive to applied flux, $\Phi_{app}$, and temperature, $T$.

Equation \ref{nondim} is also linear in the high power limit (where it is assumed $\delta_{rf}>>1$ since $\beta_{rf}\sin{\delta}+\delta \approx \delta$ for $\beta_{rf}<1$), where it separates into the following time dependent and time independent equations:
\begin{equation}
2 \pi f_{DC}= \delta_{DC}
\end{equation}
\begin{equation}
\label{highrfdiff}
2 \pi f_{rf} \sin{(\Omega \tau)}= \delta_{rf}+\frac{1}{Q_{geo}}\frac{d \delta_{rf}}{d \tau}+\frac{d^2 \delta_{rf}}{d\tau^2}
\end{equation}
Equation \ref{highrfdiff} has the same analytic solution as Eq. \ref{lowrfsol} with $\alpha$ replaced by 1, so unlike Eq. \ref{lowrfsol} it has no dependence on temperature or DC flux. In the high power limit the resonant frequency reduces to the same value as in the high temperature limit, Eq. \ref{geofreq}. 

\section{Design and Fabrication}
We have designed and measured both a single RF SQUID meta-atom and a dense 27x27 array of meta-atoms, which were manufactured using the Hypres 0.3 $\mu $A$/ \mu $m$^2$ Nb/AlO$_x$/Nb junction process on silicon substrates. However, measurements of other junctions from this run suggest the critical current density is closer to 0.2 $\mu $A$/ \mu$m$^2$ at 4.2 K. The superconducting loop is composed of two Nb films (135 nm and 300 nm thick) which are connected by a via and a Nb/AlO$_x$/Nb Josephson junction (see Fig. \ref{figure1}(b)). There is additional capacitance where these layers overlap (with SiO$_2$ dielectric) which is necessary to bring the resonant frequency within the measurable range. When designing the SQUIDs we have control over the inner and outer radius of the loop and thus the loop inductance $L$, the critical current of the junction $I_c$, and the overlap capacitance $C$. Values for $L$, $I_c$, and $C$ were chosen to maximize tunability within the measurable frequency range 6.5-22 GHz (dictated by the available waveguides) while remaining in the low noise
($\Gamma = \frac{2 \pi k_B T}{\Phi_0 I_c}<1$ and
$L_F = \frac{1}{k_B T}\left(\frac{\Phi_0}{2 \pi}\right)^2>> L$ \cite{Chesca1998})
and non-hysteretic ($\beta_{rf}=\frac{2 \pi L I_c}{\Phi_0}<1$) limits.

For the single meta-atom shown in Fig. 1(b), the inner and outer radii of the SQUID loop are 100 $\mu$m and 400 $\mu$m respectively with 3 $\mu$m diameter holes in the film every 10 $\mu$m to pin vortices \cite{Bermon1983}. We estimate $L$ to be 0.33 nH based on FastHenry calculations \cite{Whiteley}. The area of the Josephson junction and the area of the capacitor are 5.3 $\mu$m$^2$ and 600 $\mu$m$^2$ respectively,  yielding nominal design values of $I_c(4.2$ K$)=0.97$ $\mu$A and $C=0.32$ pF.

Figure \ref{figure1}(c) shows a portion of a 27x27 array of nominally identical RF SQUID meta-atoms with a center to center separation of 83 $\mu$m. The inner and outer radii of the SQUID loops are 30 $\mu$m and 40 $\mu$m respectively. We estimate $L$ to be 0.12 nH. The area of the Josephson junction and the area of the capacitor are 13 $\mu$m$^2$ and 1800 $\mu$m$^2$ respectively, yielding nominal design values of $I_c$(4.2 K) = 3.7 $\mu$A and C = 0.84 pF.

\section{Experiment}
The single RF SQUID (and later the 27x27 RF SQUID array) is oriented in a 7.6 cm long Ku (K) rectangular waveguide which has a single propagating mode operating frequency range from 9.5-19 GHz (15-26 GHz), as shown in Fig. \ref{figure1}(a). An attenuated microwave signal from an Agilent E8364C network analyzer excites a TE$_{10}$ mode in the waveguide producing an rf magnetic field, $B_{rf}$, perpendicular to the plane of the SQUID. The transmission, $S_{21}$, is measured after a cryogenic low noise amplifier (LNA) and a room-temperature (RT) amplifier. The experiment is conducted in a three-stage pulsed-tube cryostat with a sample base temperature of 6.5 K. The sample is magnetically shielded by a mu-metal cylinder and a superconducting niobium open cylinder inside the cryostat. Superconducting coils surrounding the waveguide bias the SQUID by generating a perpendicular DC magnetic field, $B_{DC}$ in the range from 0 to $\pm$1 $\mu$T.

The resonance is detected as a dip in the frequency dependent transmission magnitude through the waveguide $|S_{21}(\omega)|$. We have considered two methods for calculating scattering parameters from the solution for $\delta(t)$. One method is to consider the SQUID an effective medium with an effective relative permeability, $\mu_r$ \cite{Smith2006} \cite{Jung2013},
\begin{equation}
\label{mu}
\mu_r=1+F\left(\left<\frac{\Phi_{ac}}{\Phi_{rf}\sin{\omega t}}\right>-1\right),
\end{equation}
where $\Phi_{ac}$ is the AC flux response of the loop, the angle brackets represent time averaging, and $F$ is the filling fraction of the SQUID in the medium \cite{Lazarides2007} \cite{Du2008}. $S_{21}$ is proportional to the ratio of the transmitted electric field, $E_T$, to the incident field $E_0$, and can be calculated using the continuity of E and H fields at the boundaries of the effective medium and the empty waveguide as
\begin{equation}
S_{21}=\sqrt{\gamma} \frac{E_T}{E_0}=\frac{\sqrt{\gamma}}{\cos{kl}-\frac{i}{2}\left(\frac{1}{\gamma}+\gamma\right)\sin{kl}},
\label{effectivemed}
\end{equation}
where $l$ is the length of the medium, $\gamma=\frac{k}{\mu_r k_0}$, $k=\sqrt{\mu_r\left(\frac{\omega}{c}\right)^2-\left(\frac{\pi}{a}\right)^2}$ is the wave number in the medium, and $k_0=k(\mu_r=1)$ is the wave number in the empty waveguide. For the single SQUID (as opposed to the array) the choice of $l$ and $F$ is not straightforward. They are used as fitting parameters (along with resistance $R$) for the width and depth of the measured resonances. The important parameter for the fit is the relationship between $F$ and $l$, which we find to be $F=\frac{A_{loop}}{0.79 l}$, where $A_{loop}$ is the area of the SQUID loop (calculated from the average radius).

An alternate method for estimating $S_{21}$ focuses on the power dissipated in the resistor, $R$ in Fig. \ref{figure1}(d).
\begin{equation}
S_{21}=\sqrt{\frac{P_T}{P_0}}=\sqrt{1-\frac{V^2/R}{P_0}}
\label{disspower}
\end{equation}
where $P_T$ is the transmitted power and $P_0$ is the incident power. This assumes that the only power not transmitted through the waveguide is the power dissipated in the resistor. It does not account for reflection or other loss mechanisms.

\section{Results}
\label{results}
\begin{figure}[]
\includegraphics*[width=80mm]{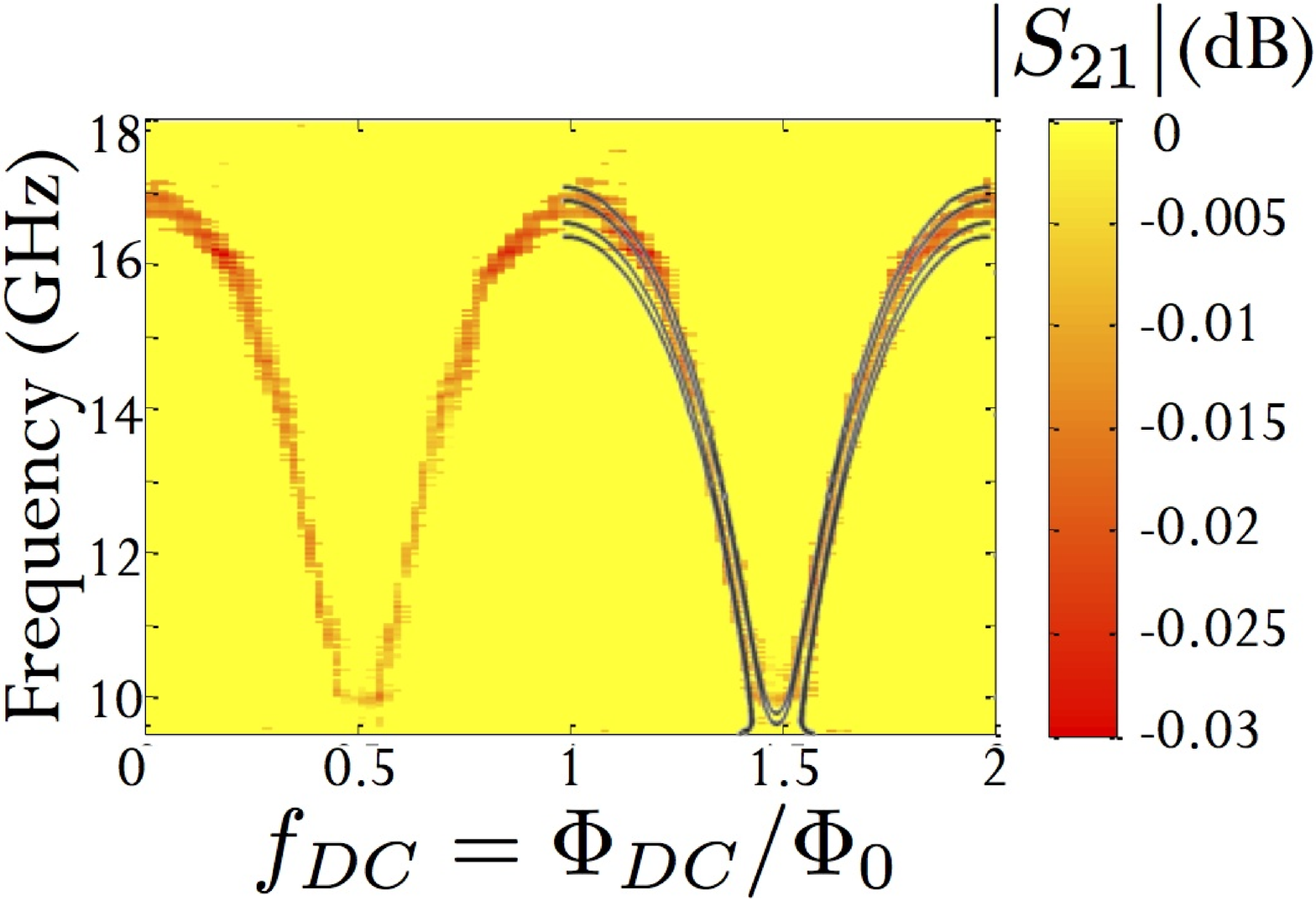}
\caption{$|S_{21}|$ of a single RF SQUID meta-atom as a function of frequency and applied DC flux at -80 dBm rf power and 6.5 K. The resonant response is identified by the red features. There are contour lines for $|S_{21}|$ at -.005 and -.01 dB generated by the model, Eq. \ref{effectivemed}.}
\label{fluxsweep}
\end{figure}

The resonance of an RF SQUID responds to three tuning parameters: DC magnetic field, temperature, and rf power. Modifying the resonance with DC magnetic field is the most straightforward. Figure \ref{fluxsweep} shows the experimental results for $|S_{21}|$ of an RF SQUID meta-atom as a function of DC flux and frequency.
The incident rf power and the temperature were fixed at -80 dBm and 6.5 K, respectively. Resonance dips in $|S_{21}(\omega)|$ appear as the red features against a yellow background of unaffected signals ($|S_{21}|=0$dB). The signal was extracted by subtracting $|S_{21}|$ at 16 K (which is well above the critical temperature) from $|S_{21}|$ at 6.5 K, removing a background variation, and applying a threshold to identify the resonance. The resonance shows good periodicity with DC flux, with a maximum value of $16.9\pm0.3$ GHz and a minimum at or below 10 GHz. The cutoff frequency of the Ku waveguide is 9.5 GHz which imposes a lower frequency limit on this measurement. The same sample measured in an X band waveguide (single propagating mode operating frequency range from 6.6-13 GHz) has a minimum resonant frequency of $9.5\pm0.5$ GHz. 

The small magnitude of the resonance dips can be attributed to the small size of the SQUID relative to the waveguide. This becomes less of a factor when considering arrays of SQUIDs which occupy a more appreciable fraction of the waveguide cross section. The quality factor, $Q$, is extracted by fitting the $|S_{21}(\omega)|$ data to a model of an RLC-resonator inductively coupled to a transmission line \cite{Doyle2008}. $Q$ also shows periodicity with DC flux, following the same trend as the resonant frequency (Eq. \ref{Q}), with a maximum value of $54$ and minimum at or below $20$.

At an rf power of -80 dBm, ($\Phi_{rf} \approx 0.003\Phi_{0}$) the low power linearized equation (Eq. \ref{lowrfdiff}) is valid (since $\delta_{rf}<<1$). 
Treating the RF SQUID as an effective medium in the waveguide, $|S_{21}(\omega, \Phi_{DC})|$ is calculated from Eqs. \ref{dconly}, \ref{lowrfsol}, \ref{mu}, and \ref{effectivemed} and plotted as the contour lines in Fig. \ref{fluxsweep}.
The capacitance $C$ and critical current $I_c$(6.5 K) are adjusted from their nominal values and taken to be 0.42 pF and 0.55 $\mu$A,  respectively. These values were chosen so the resonant frequency at $T=6.5$ K and $\Phi_{DC}=0$ agree with the measurements in both the low and high power limits i.e. $f_0(6.5$ K$, \Phi_{DC}=0)=16.8$ GHz for low power (Eq. \ref{resfreq}) and $f_{0,geo}=13.5$ GHz (Eq. \ref{geofreq}) for high power.
The model and the data agree both in terms of the flux dependence and magnitude of the tunability ($\sim7$ GHz), and the magnitude of the resonance dip.

\begin{figure}[]
\includegraphics*[width=80mm]{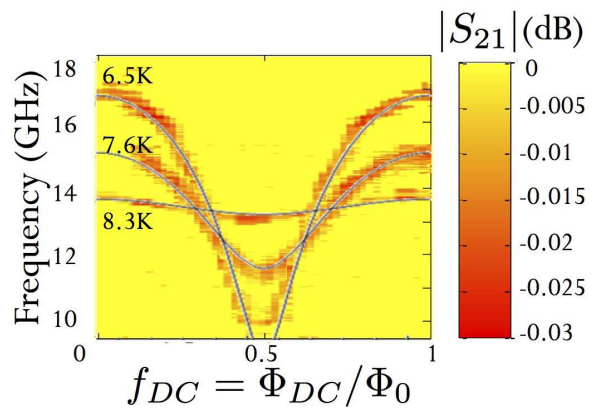}
\caption{$|S_{21}|$ of a single RF SQUID meta-atom as a function of frequency and applied DC flux at three temperatures, 6.5 K, 7.6 K, and 8.3 K, and -80 dBm rf power. The resonant response is identified by the red features. The solid lines are the resonant frequency calculated by Eqs. \ref{Ljj}, \ref{resfreq}, and \ref{dconly}.}
\label{tdepend}
\end{figure}

The DC flux tunability is modified by changes in temperature.
The red features in Fig. \ref{tdepend} show the DC flux tuned resonance at 6.5 K, 7.6 K, and 8.3 K for an rf power of -80 dBm.
The flux tunability is reduced from $\sim7$ GHz at 6.5 K, to $\sim3$ GHz at 7.6 K, and $\sim1$ GHz at 8.3 K. 
The quality factor, $Q$, is 50 at $\Phi_{DC}=0$ and does not show significant variation with temperature.
The grey lines follow the resonance by using the solution for $\delta_{DC}$ from Eq. \ref{dconly} and calculating the resonant frequency using Eqs. \ref{Ljj} and \ref{resfreq} (assuming $\delta \approx \delta_{DC}$). The reduction in tunability with increased temperature in consistent with the model. The only parameter varied to fit this data is the temperature dependent critical current. 

We also measured $|S_{21}|$ as a function of frequency and dc flux at low power for various fixed temperatures ranging from 6.5 to 9.2 K (not shown). For $f_{DC}=0$ the resonant frequency decreased from 17 to 13.5 GHz with increasing temperature and for $f_{DC}=1/2$ it increased from 9.5 to 13.5 GHz over the same temperature range. Consistent with the model, the resonant frequency saturates in the high temperature limit at $f_{0,geo}=13.5\pm0.2$ GHz (Eq. \ref{geofreq}).

\begin{figure}[]
\includegraphics*[width=80mm]{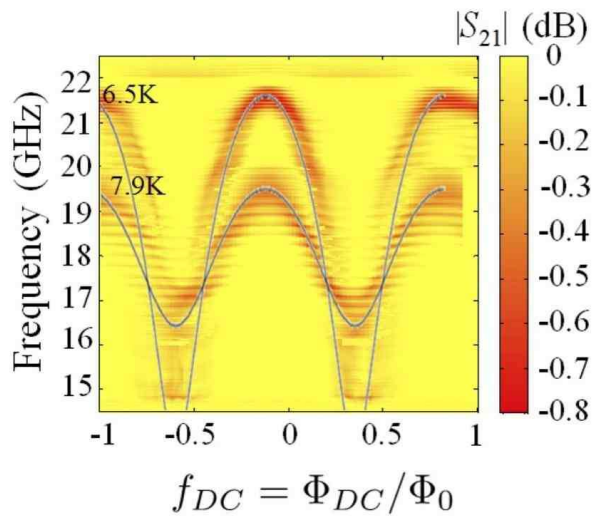}
\caption{$|S_{21}|$ of the 27x27 RF SQUID array as a function of frequency and applied DC flux at two temperatures, 6.5 K and 7.9 K, and -$60$ dBm rf power. The resonant response is identified by the red features. The solid lines are the resonant frequency calculated for a single SQUID by Eqs. \ref{Ljj}, \ref{resfreq}, and \ref{dconly}.}
\label{array}
\end{figure}

The temperature dependence of DC flux tuning for the 27x27 RF-SQUID array is shown in Fig. \ref{array}.
The resonance (red features) shows good periodicity with DC flux with a maximum of $21.5\pm0.2$ GHz and a minimum below $15$ GHz at 6.5 K and -$60$ dBm rf power. As temperature increases, the flux tunability is reduced from $> 6.5$ GHz for 6.5 K to $2.5$ GHz at 7.9 K. A flux offset of about $0.1 \Phi_{0}$ shifts the curve along the flux axis without affecting the periodicity. The periodic horizontal features result from standing waves in the system due to an impedance mismatch in the external circuit.

The meta-atoms in the array respond to the tuning parameters coherently, creating a stronger signal and a slightly wider resonance dip compared to a single RF SQUID. The flux dependence of the array is qualitatively the same as that of a single SQUID. This is demonstrated by the grey lines in Fig. \ref{array} which are solutions for the resonant frequency of a single SQUID with the following parameters: $L=0.12$ nH (nominal value), $C=0.65$ pF, $I_c$(6.5K)$=1.2$ $\mu$A, and $I_c$(7.9K)$=0.47$ $\mu$A.

\begin{figure}[]
\includegraphics*[width=80mm]{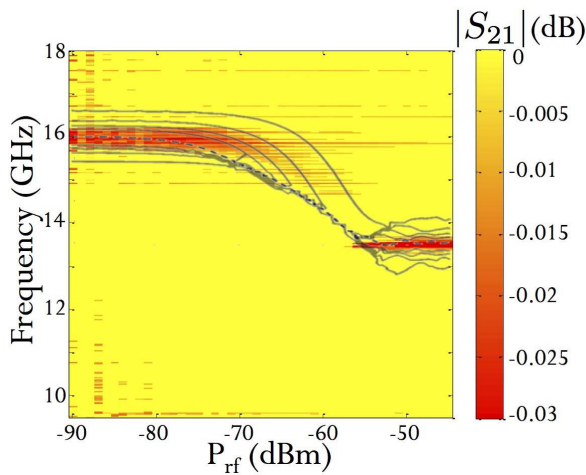}
\caption{$|S_{21}|$ of a single SQUID meta-atom as a function of frequency and rf power at fixed DC flux, $f_{DC}=1/6$ and temperature $T=6.5$ K. The resonant response is identified by the red features. The contour lines for $|S_{21}|$ at -.04, -.03, -.02, and -.01 dB are generated by the model, Eq. \ref{disspower}. The dashed line indicates the model-determined resonant frequency (minimum of $|S_{21}(\omega)|$).}
\label{rfpower}
\end{figure}

The resonance of an RF SQUID can also be modified by current induced by $\Phi_{rf}$. Figure \ref{rfpower} shows $|S_{21}|$ as a function of rf power and frequency for a single RF SQUID meta-atom in the representative case of $f_{DC}=1/6$ and $T=6.5$ K.
The resonant response is represented by the red features. When the rf power is low, the resonant frequency remains at a fixed value, 16 GHz, and the resonance dip has a constant depth. As the rf power increases, the resonant frequency decreases and the resonance dip becomes shallower. The resonance disappears at intermediate powers before reappearing as a strong resonance at high power (-55 dBm). 

Calculating the rf power dependence of the resonance requires a full numerical solution of the nonlinear Eq. \ref{nondim}. Instead of regarding the RF SQUID as an effective medium, we employ Eq. \ref{disspower} to estimate $|S_{21}|$. The results of this calculation are plotted as contour lines in Fig. \ref{rfpower} and the calculated minimum of $|S_{21}|$ is plotted as a dashed line. Both show excellent agreement with the data, including the depth of the $|S_{21}|$ dip.

There are two transition points in the rf power dependent resonance at $-75$dBm ($\Phi_{rf} \approx 0.006 \Phi_0$) and -57 dBm ($\Phi_{rf} \approx 0.05 \Phi_0$). We can understand the significance of these powers and the behavior in the low, intermediate, and high power regions using the model discussed in Sec. \ref{model}. In the low power limit, the resonance does not change with increasing rf power because $S_{21}$ from Eq. \ref{disspower} does not depend on input rf power, $P_0$, since $P_0 \propto V^2$ ($V \propto f_{rf}$ from Eq. \ref{lowrfsol} and $f_{rf} \propto \sqrt{P_0}$ because of the relationship between input power and magnetic fields in a waveguide).

The first transition occurs at an rf power of $-75$ dBm, where the resonant frequency begins to decrease and the resonance dip becomes shallower. At this rf power the oscillation amplitude of $\delta$ during an rf period is $\pi/2$ on resonance. The resonant frequency is modified because above $-75$ dBm rf power, the time averaged phase difference $\left< \delta \right>$ deviates from $\delta_{DC}$. In the case of $f_{DC}=1/6$, $\left< \delta \right> > \delta_{DC}$ resulting in a decrease in $\cos{\delta}$ from its low power value, a corresponding increase in $L_{JJ}$ (Eq. \ref{Ljj}), and an ultimate decrease in resonant frequency $f_0$ (Eq. \ref{resfreq}). However in the case of $f_{DC}=1/2$, the modification in $\left< \delta \right>$ causes an increase in $\cos{\delta}$ from its low power value and an ultimate increase in resonant frequency, $f_0$. This is consistent with the experimental results (not shown). The resonance dip becomes shallower because $V$ is no longer proportion to $f_{rf}$; instead the input power, $P_0$, increases more than the dissipated power, $V^2/R$.

The second transition occurs at an rf power of $-57$ dBm where $\delta$ oscillates with an amplitude exceeding $2\pi$ on resonance and the Josephson junction undergoes multiple phase slips in each rf period. This is the high rf power limit (Eq. \ref{highrfdiff}), where the tunable resonant frequency reduces to a fixed value (Eq. \ref{geofreq}), i.e. $f_{0,geo}=13.5$ GHz. We expect the losses to be greater in this regime because the phase slips are dissipative, resulting in a deeper resonance dip, which is clearly evident in the data.

A similar evolution of a transition frequency with RF power has been observed in single qubits coupled to microwave cavities containing a small number of photons \cite{Reed2010}\cite{Novikov2013}.  The Janes-Cummings Hamiltonian shows such behavior at high photon excitation number, reproducing the observed dispersion and amplitude variation of the transition with increasing RF power \cite{Bishop2010}.

\section{Discussion}
\label{discussion}
The design parameters of the SQUID meta-atom are the critical current $I_c(T)$, the capacitance $C$, the sub-gap resistance $R$, and the loop inductance $L$.  In our comparison of data and model for the single meta-atom we use the nominal design value for loop inductance because the lithographic process dictates the loop geometry to high precision. We adjusted the capacitance $C$ and critical current $I_c$(6.5 K) from their nominal design values to fit the experimentally observed resonant frequencies at $P = -80$ dBm using Eq. \ref{resfreq} and $P=-50$ dBm using Eq. \ref{geofreq} for $T = 6.5$ K and $\Phi_{DC} = 0$. The fit value for capacitance is 30$\%$ higher than the design value, which exceeds the 20$\%$ tolerance quoted for the Nb/AlO$_x$/Nb process \cite{HypresRules}. (The same method was used to determine $I_c$(6.5 K), $I_c$(7.9 K), and $C$ for the fit to the array data in Fig. \ref{array}.) The sub-gap resistance $R=820$ $\Omega$, and the filling fraction $F$ and medium length $l$ in Eqs. \ref{mu} and \ref{effectivemed} were adjusted to match the width and depth of the observed $|S_{21}(\omega)|$ minimum at $\Phi_{DC} = 0$, $T = 6.5$ K, and -80 dBm. With these parameters we quantitatively explain the DC flux and rf power dependence of the RF SQUID meta-atom.

\begin{figure}[]
\includegraphics*[width=80mm]{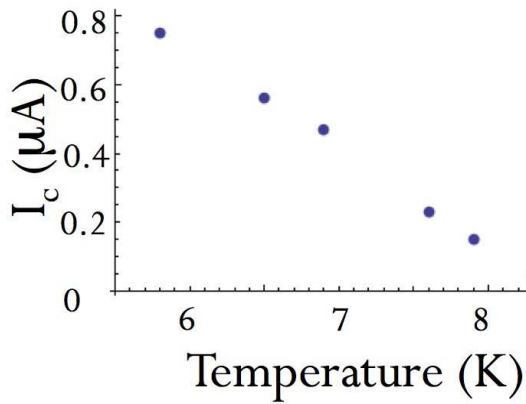}
\caption{Critical current temperature dependence $I_c(T)$ calculated from measured resonant frequencies for a single SQUID at zero DC flux and corresponding fixed temperatures}
\label{tempfit}
\end{figure}

We extracted the temperature dependence of the critical current $I_c(T)$ from the flux dependence of the resonant frequency at different fixed temperatures. The critical current of the junction at each temperature was calculated by substituting the maximum resonant frequency as a function of flux into Eqs. \ref{Ljj} and \ref{resfreq}. The results for $I_c(T)$, shown in Fig. \ref{tempfit}, are consistent with previous results on Nb/AlO$_x$/Al/Nb tunnel junctions \cite{Nevirkovets2001}.

We have demonstrated tunability of the RF SQUID meta-atom with temperature, RF current, and DC magnetic flux. The SQUID meta-atom has an unusual response to changes in temperature. The resonant frequency can either increase or decrease, depending on the DC magnetic flux applied to the meta-atom; the slope of temperature tuning of $f_0$ can be large or small, and either positive or negative.

Tuning with rf power can also either increase or decrease the resonant frequency, depending on the applied DC magnetic flux. However, the strength of the resonance response varies with rf power while it remains constant with temperature tuning. At low powers ($<-75$ dBm), and high powers ($>-57$ dBm), the meta-atom has a resonant interaction with electromagnetic radiation, but at intermediate powers it becomes essentially transparent.  This is similar to demonstrations of metamaterial-induced transparency in which a tunable ``spectral hole" is created by interfering resonant processes in two or more meta-atoms making up a meta-molecule \cite{Fedotov2007} \cite{Papasimakis2008} \cite{Tassin2009} \cite{Kurter20113}. There, the transparency condition in the superconducting meta-molecule could be suppressed in a switching transition at high power \cite{Kurter2012}. In our case, the transparency process is due to the non-linear Josephson effect and is self-induced, making transparency in the RF SQUID meta-atom simpler than previous implementations. Also unlike other realizations, the transparency condition arises from a decrease in dissipation without enhancement in loss at nearby frequencies \cite{Abdumalikov2010}. As such it is potentially useful as a power limiter for sensitive front-end receivers \cite{Mukhanov2004}.

The DC flux tuning of the SQUID meta-atom is remarkably sensitive. At its maximum value, the flux tunability (defined as the frequency change divided by the change in magnetic field) is approximately 80 THz/Gauss, at 12 GHz and 6.5 K.  Hence only extremely small magnetic fields are required to make very substantial changes in the properties of the meta-atom.

The next question is how quickly can tuning by each parameter be achieved. The shortest time scale for tuning a superconductor is $\sim \hbar/\Delta\sim 1$ ps, where $\Delta$ is the energy gap. In this case, the RC time constant of $0.3$ ns (L/R time constant $0.4$ ps) imposes an upper limit on the intrisic switching speed. Tuning with temperature depends on changes to the critical current of the junction and is relatively slow, with typical estimates on the order of 10 $\mu$s \cite{Savinov2012}.
A pulsed rf power measurement of the SQUID meta-atom is consistent with a response time less than $500$ ns.  Hence the rf power tuning time is in the sub-$\mu$s range perhaps only limited by the RC time. Flux tuning of SQUID-like superconducting qubits has been accomplished on nano-second time scales limited only by the rise-time of the applied current pulse \cite{Paauw2009} \cite{Zhu2010}. 

The tunability of the meta-atom in normalized units (defined as tuning frequency range divided by the center frequency of tuning) is 56$\%$.  The meta-atom loss varies by about 64$\%$ (defined as the loss range divided by maximum loss) over this entire range. We estimate the figure of merit (defined as the ratio of the real part of $\mu$ to the imaginary part) to be 1700 on resonance in the case of low power and zero DC flux. The goal of achieving wide meta-atom tunability on short time scales with low loss is well within reach.

The coherent response of the meta-atoms in the 27x27 array demonstrates the feasiblity of a widely tunable RF SQUID metamaterial. The SQUIDs are close enough that their interactions play a role in their dynamics. We define a coupling parameter which is the ratio of mutual inductance to self-inductance; $\kappa\cong\frac{\mu_0 \pi r^4}{4 d^3 L}=0.02$, where $r$ is the radius of the loop and $d$ is the center-to-center spacing of the SQUIDs.
The ratio of the wavelength to the lattice parameter for the measured array is $\sim 350$, which is well-within the meta-material limit. The array displays an overall tunability of $46\%$ and a flux tunability of 4 THz/Gauss at 19 GHz and 6.5K. Future work will further explore arrays of RF SQUID meta-atoms in both the interacting and non-interacting limits.

\section{Conclusion}
\label{conclusion}
We have designed, fabricated, and presented an rf superconducting quantum interference device meta-atom in a geometry that can be scaled for free-space interactions with electromagnetic fields.  The meta-atom proves to be highly tunable with dc magnetic field, rf currents, and temperature.  The RF SQUID meta-atom proves to have low losses over its wide (56$\%$) range of tunability. A novel form of electromagnetically-induced transparency has been observed in this meta-atom with increasing induced rf current. We have obtained a detailed quantitative understanding of this nonlinear meta-atom with first principles modeling.  The data and model are in excellent agreement. Finally, a two-dimensional array of RF SQUIDs has shown coherent tuning with DC flux and temperature in a manner similar to a single SQUID.

\begin{acknowledgments}
This work is supported by the NSF-GOALI program through grant $\#$ECCS-1158644, and the Center for Nanophysics and Advanced Materials (CNAM). We thank Masoud Radparvar, Georgy Prokopenko, Jen-Hao Yeh and Tamin Tai for experimental guidance and helpful suggestions, and Alexey Ustinov, Philipp Jung and Susanne Butz for helpful discussions. We also thank H. J. Paik and M. V. Moody for use of the pulsed tube refrigerator.
\end{acknowledgments}

\bibliography{bibliography}
\end{document}